\documentclass[letterpaper, 10 pt, conference]{ieeeconf}  

\IEEEoverridecommandlockouts                              

\usepackage{amsmath} 
\usepackage{amssymb}  

\usepackage{graphicx} 
\usepackage{algorithm} 
\usepackage{algpseudocode}  
\usepackage{booktabs} 
\usepackage{listings}
\usepackage{soul} 
\usepackage[backend=bibtex]{biblatex}

\newcommand{\citet}[1]{\citeauthor{#1} (\citeyear{#1}) \cite{#1}} 
 
\ifdefined\preamble\else

\newcommand{\y}{\mathbf{y}}

\newcommand{\preamble}{}

\fi

\title{\LARGE \bf
Online Fine-Tuning of Carbon Emission Predictions using \\Real-Time Recurrent Learning for State Space Models
}

\author{Julian Lemmel$^{1,2}$, Manuel Kranzl$^{2}$, Adam Lamine$^{2}$, \\Philipp Neubauer$^{2}$, Radu Grosu$^{1}$, Sophie Neubauer$^{2}$
\thanks{$^{1}$ TU Wien, Karlsplatz 13, 1040 Vienna, Austria}%
\thanks{$^{2}$ Pyrosoma AI, Am Europlatz 5, Bldg. 1C/EG, 1200 Wien, Austria}
\thanks{{\tt\small julian.lemmel@tuwien.ac.at}}%
}

\begin{document}

\maketitle
\thispagestyle{empty}
\pagestyle{empty}

\begin{abstract}


This paper introduces a new approach for fine-tuning the predictions of structured state space models (SSMs) at inference time using real-time recurrent learning. While SSMs are known for their efficiency and long-range modeling capabilities, they are typically trained offline and remain static during deployment. Our method enables online adaptation by continuously updating model parameters in response to incoming data. We evaluate our approach for linear-recurrent-unit SSMs using a small carbon emission dataset collected from embedded automotive hardware. Experimental results show that our method consistently reduces prediction error online during inference, demonstrating its potential for dynamic, resource-constrained environments.
\end{abstract}


\section{INTRODUCTION}

Vehicle emission prediction is a critical area of research due to its implications for environmental sustainability and public health. Predicting emissions from data readily available on modern automobiles can help estimate effects on air quality and climate alike. The advent of deep learning has revolutionized emissions prediction by enabling data-driven approaches that leverage large-scale datasets from chassis dynamometer tests and on-road measurements.

Recurrent neural networks (RNNs) are capable of modeling temporal dependency in time-series data and have been successfully applied to emission prediction tasks, such as forecasting nitrogen oxides (NOx) emissions from vehicles under varying traffic conditions~\cite{zhang2018, zhong2024}. Recently, state space models (SSMs) such as LRUs~\cite{gu2021a}, have revitalized research on RNNs, as foundational large language models.

Continual learning approaches, such as incremental retraining, are essential to ensure that emission prediction models remain accurate as new data become available ~\cite{zhong2024}. Online learning techniques, which process data in real-time \cite{williams1989, lemmel2025}, are particularly relevant for applications in intelligent transportation systems and digital-twin platforms. These technologies enable models to adapt to evolving traffic conditions, vehicle technologies, and environmental factors, ensuring consistent performance over time.

In this work, we show how LRUs, pretrained on a small carbon emission data-set collected from embedded hardware, can be fine-tuned online through real-time recurrent learning \cite{williams1989}. The fine-tuning successfully refines the emission predictions. Our results are very encouraging and generalizable to fine-tuning large language models during inference time.

\section{RELATED WORK}
Traditional models, such as COPERT and EMFAC, have been widely used for macro-level analysis, providing estimates based on aggregated data \cite{ntziachristos2009, fujita2012}. However, these models often lack the granularity required for real-time applications and fail to account for dynamic driving conditions and regional variability. Modal models like MOVES and CMEM offer higher temporal resolution but rely on static assumptions, limiting their adaptability to real-world scenarios \cite{zhong2024}. 
Artificial neural networks have been widely applied to model complex relationships between vehicle attributes, driving patterns, and emissions. Advanced architectures, such as ST-MFGCN\cite{xu2021}, further enhance spatial-temporal correlations by integrating graph-based learning and attention mechanisms. These models address the limitations of traditional methods by dynamically adapting to spatial and temporal variations in emission data \cite{zhong2024}.

RNNs are used for modeling temporal dependencies in the data. They are different from classical feedforward neural networks due to recurrent connections resulting in a hidden state that is maintained across entire trajectories. These models are particularly effective in scenarios where emissions are influenced by sequential driving behaviors, such as acceleration, deceleration, and idling \cite{zhong2024, mahendra2024}. For instance, \parencite{zhang2018} demonstrated the effectiveness of LSTMs in predicting vehicle emissions by integrating weather variables, driving patterns, and historical emissions data \cite{zhang2018}. Similarly, GRUs have shown efficiency in handling shorter forecasting horizons, making them suitable for real-time applications \cite{mahendra2024}.

To the best of our knowledge, none of the works discussed above has addressed the problem of online fine-tuning the carbon-emission predictions of SSMs, in particular of LRUs, through real-time recurrent learning.  

\section{DATASET}

We empirically test our approach on a small dataset that was recorded from specialized embedded sensor hardware developed by the Edge Intelligence Research Institute Nanjing. The dataset consists of a total of 34244 samples and 12 columns. The columns are:
%
(1)~Timestamp (s),
(2)~Engine speed rpm (rounds/min),
(3)~Hourly fuel consumption (litre/hour),
(4)~Coolant temperature (°C),
(5)~Speed (kilometre/hour),
(6)~Instantaneous fuel economy (kilometre/litre),
(7)~NO ppm (parts per million),
(8)~NO2 ppm,
(9)~NOX ppm,
(10)~CO2, and
(11)~CO ppm.

The dataset comprises 5 recording sessions with a total of 12:02:58 of recorded time. The median time delta between consecutive data points on any given day was 1.0 second, while the mean time delta was \textasciitilde1.267 seconds. This implies that a
considerable number of measurements are missing. 
\begin{figure*}
    \centering
    \includegraphics[width=\linewidth]{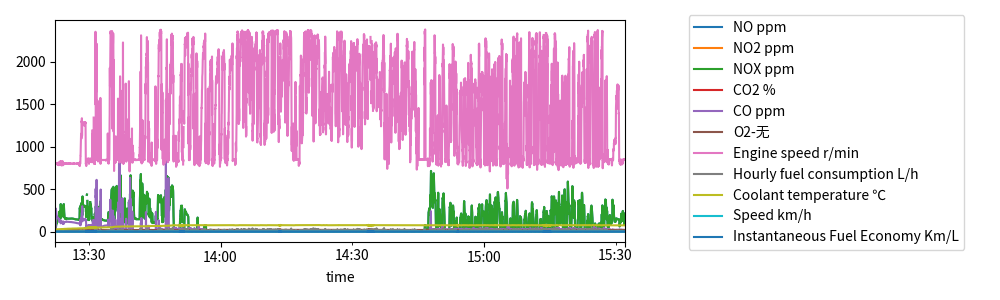}
    \caption{Plot of the raw data for one recorded drive.}
    \label{fig:rawdata}
\end{figure*}
For our carbon emission prediction problem, the last five columns form the target values, while the remaining columns are the explaining variables. For fitting a model, we selected the first 4 days as our training set and the remaining days as the validation set. This corresponds to a 80/20 split that is commonly found in the literature. Fig. \ref{fig:rawdata} shows the raw data of the recording used for validation.

Automobile carbon emissions naturally depend on weather conditions. For
example, low and high temperatures may require the use of heating and
air conditioning, respectively, both leading to an increase in power
demand and subsequently carbon emissions. In order to capture the
dependency of carbon emissions on the current weather, we acquired Nanjing regional weather data for the days of recording from
\url{https://www.visualcrossing.com/}. It includes hourly measurements
for a range of indicators such as temperature, precipitation, or a categorical description of the weather. The weather data was concatenated with the emission data by matching the timestamp.

\section{METHODS}
Carbon emission prediction is a time-series prediction task that can be naturally approached using sequence models such as RNNs. For this work, we implemented a carbon emission model using SSMs, and later fine-tuned them during inference using real-time recurrent learning (RTRL) \cite{williams1989,lemmel2025}. In the following, we will first introduce SSMs and RTRL. We will give an intuition as to why SSMs have a strong computational advantage when trained using BPTT, and also how they allow for computing the exact gradient with reasonable computational complexity when using RTRL.

\subsection{Sequence Models}
When doing Machine Learning on sequential data it is highly beneficial for the model to retain a memory of preceding inputs. Sequence models such as Recurrent Neural Networks (RNNs) can do just that by persisting a hidden state that is used when computing future outputs. Due to the evolving hidden state ("memory"), they exhibit temporal dynamic behavior which enables them to accurately model temporal dependencies in time-series data. RNNs are traditionally trained using backpropagation through time (BPTT), where an input sequence of fixed length is fed to the network, producing a corresponding sequence of outputs. The outputs are used for computing the loss and its gradient where inputs at the beginning of the sequence influence all succeeding outputs. In the following we introduce two instantiations of RNNs that were used for this project: Linear Recurrent Units and Continuous-Time RNNs.

\subsection{Linear State Space Models}
State-Space Models (SSMs) have recently been used for tasks requiring long-term dependencies and sequence modeling due to their ability to handle complex temporal dynamics \cite{gu2021a}. SSMs are a RNNs resulting from discretizing complex-valued dynamical systems, described by a Linear ODE:
\begin{equation}
    \begin{aligned}\frac{d}{d t} h_{\mathrm{ct}}(t)&=A\ h_{\mathrm{ct}}(t)+B\ u_{\mathrm{ct}}(t) \\y_{\mathrm{ct}}(t)&=\Re\left[C\ h_{\mathrm{ct}}(t)\right]+D\ u_{\mathrm{ct}}(t)\end{aligned}
\end{equation}
Linear Recurrent Units (LRUs) are State Space Models that impress with comparably simple architecture and initialization. The seminal work by \citet{orvieto2023} has been pivotal in advancing the understanding and application of LRUs. In a comprehensive ablation study, they demonstrated that linearizing, and diagonalizing the recurrence, normalization, and using stable exponential parameterization were imperative for the performance of SSMs on long-range reasoning tasks, but not the initialization of parameters using the HiPPO theory central to preceding SSMs such as S4 \cite{gu2021a}. LRUs were shown to achieve performance on par with S4, while also matching their computational efficiency and making use of a much simpler initialization scheme. LRUs are complex-valued linear RNNs with diagonal connectivity that result from discretizing the ODE seen before.
\begin{equation}
    \begin{aligned} h_{t+1}&=\Lambda\ h_{t}+B\ u_{t+1} \\y_{t}&=\Re\left[C\ h_{t}\right]+D\ u_{t}\end{aligned}
\end{equation}
Here, diagonal connectivity means that the connectivity matrix $\Lambda$ is a diagonal matrix. This attribute enables SSM state-updates to be computed in parallel using a parallel scan \cite{BlellochTR90}. Since each node's state does not depend on any parameters belonging to different nodes, it further allows Real-Time Recurrent Learning (RTRL) updates to be computed efficiently, as shown in \citet{zucchet2023}.

\subsection{Real-Time Recurrent Learning}
An RNN is a function $f$ applied to sequences that receives an input $x_{t}$ at time $t$ and a hidden state $h_{t}$ and produces an output $\hat y_t$ and the next hidden state $h_{t+1}$. The function $f$ also receives learnable parameters $\theta_{t}$.
\begin{equation}
    \hat y_{t}, h_{t+1} = f(x_{t}, h_{t},\theta_{t})
\end{equation}
Real Time Recurrent Learning (RTRL) \cite{williams1989,lemmel2025} is a method for training RNNs in which updates are computed \emph{online}. This is made possible by estimating the Jacobian of the RNN update function. The estimation is updated alongside the hidden state at each step and used to map downstream error-vectors to parameter gradients.  
\begin{equation}
    \hat y_{t}, h_{t+1}, J_{t+1} = f(x_{t}, h_{t}, J_{t}, \theta_{t})
\end{equation}
The update rules used in RTRL are derived in the following. Given a dataset consisting of multivariate time-series $x(t)\in \mathbb R^I$ of inputs and $y(t) \in \mathbb R^O$ of labels we want to minimize some Loss function  $\mathcal L_{\theta}=\sum_{t=0}^{T} L_{\theta}(y_{t})$ by gradient descent. This is achieved by taking small steps in the direction of the negative gradient of the total loss. 
\begin{equation}
    \Delta \theta =-\eta \nabla_{\theta}\mathcal L_\theta = -\eta\sum\limits_{t=0}^{T}\Delta \theta(t)=-\eta\sum_{t=0}^{T} \nabla L_{\theta}(y_{t})
\end{equation}
We can also compute the gradient of the loss at each step.
\begin{equation}
    \Delta \theta(t)=\nabla_{\theta}L_{\theta}(t) = \nabla_{\theta}\hat y_{t}\nabla_{\hat y_{t}}L_\theta(t)
\end{equation}
with $\hat y_{t}$ being the output of the RNN at timestep $t$.

RNNs maintain a hidden state, which is influenced by the same set of parameters, at each step in the sequence. This must be taken into account when computing the gradient of the loss. While BPTT unrolls the recurrence a fixed number of times before computing the gradient, RTRL keeps the recurrence's Jacobian $J_{t}$ (defined as the total derivative $J_t\,{=}\,dh_t{/}d\theta$ of the hidden state with respect to the parameters) as an eligibility trace $J_{t}$, that is updated at each step. The trace (Jacobian) $J_{t}$ is computed iteratively, by linearly combining the \textit{immediate} Jacobian $\partial h_{t}/\partial\theta$, with the previous trace $J_{t-1}$ scaled by the recurrent influence $\partial h_{t}/\partial h_{t-1}$:
\begin{equation}
    \quad J_{t}=J_{t-1}\ \frac{\partial h_{t}}{\partial h_{t-1}} + \frac{\partial h_{t}}{\partial\theta}
\end{equation}
This equation allows the forward computation of the Jacobian, in parallel to the computation of the hidden state and output of the RNN. When taking an optimization step, we can then calculate the actual gradients by combining with the downstream gradient as $J_t  \nabla_{\hat y_{t}}L_{\theta_t}$. And subsequently the parameters can be updated with $\quad \theta \leftarrow \theta - \eta\ J_t  \nabla_{\hat y_{t}}L_{\theta_t}$

RNNs are traditionally trained using BPTT, where an input sequence of fixed length is fed to the network, producing a corresponding sequence of outputs. The outputs are used for computing the loss and its gradient, where inputs at the beginning of the sequence influence all succeeding outputs. 

RTRL update RNN's parameters \textit{online}, after each input, as discussed above.
%
%
For arbitrary recurrent weight matrices, however, RTRL has a worse computational complexity than BPTT and as a consequence is not used much. 

However, the SSMs discretization ensures diagonal connectivity in the recurrent layer. In other words, the connectivity matrix $\tilde A$ is diagonal. Stated differently, a node's state depends only on itself and the input. This enables efficient computation of the Jacobian-trace updates, that is constant in the number of nodes \cite{zucchet2023}. Intuitively, $\partial h_t/\partial h_{t-1}$ is a diagonal matrix, and therefore the term simplifies to a vector-matrix product. 
%
LRUs trained with RTRL can be generalized to a multi-layer network setting, and therefore LRU-RNNs are a viable option for efficient online learning.

\subsection{Preprocessing}
The overall data-preprocessing pipeline involves the following steps,
which are described in detail in the remainder of this section:
%
(1) One-hot encode categorical columns,
(2) Standardize numerical columns,
(3) Impute missing values (NaNs), and
(4) Apply augmentation to create synthetic data.
%
\begin{figure}
\centering
\includegraphics[width=\linewidth]{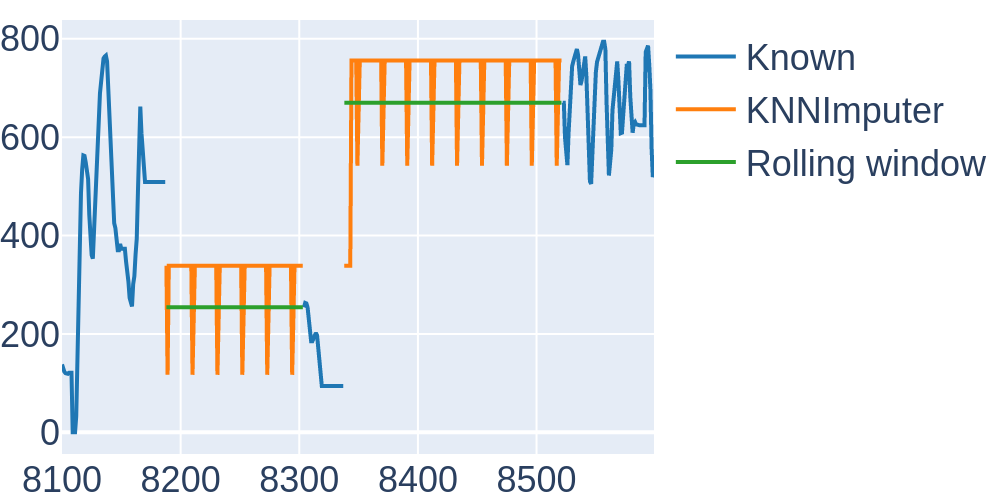}
\caption{The KNNImputer found by cross-validation, vs.~our naive rolling window approach. The KNNImputer introduces unnatural looking data.}\label{fig:dcm_imputation}
\end{figure}
\paragraph{One-hot Encoding}
Since most time-series-related machine-learning techniques can only
handle numerical values, we had to encode categorical values in a way
that is compatible with these methods. We chose to address this using one-hot encoding, meaning that each categorical column is expanded into as many columns as there are distinct values within the dataset. For each row, all new columns are set to zero except the
one corresponding to the original value, which is set to one. This was done to both training and validation set simultaneously to ensure the same number of feature dimensions during training and validation. 

\paragraph{Standardization}
The majority of machine learning algorithms assume data with zero mean
and unit variance. Therefore, we standardized all columns by subtracting the mean
and dividing by the scale. Here, we used the statistics of the training
set for both training and validation.

\paragraph{Missing Data Imputation}
For our first naive implementation,~ we
chose KNN-imputation \cite{murti2019}, a technique that leverages
nearest neighbors to impute the data-points in question. Specifically,
the missing values are computed as the mean of the K nearest neighbors
that have a value in the column of interest, where neighborhood is
computed from the non-missing values.
The best KNNImputer found using scikit-learn cross-validation was
with n\_neighbors=20 and weights="uniform". We found that the imputation results did not look good qualitatively: large jumps were introduces that led to overall unnatural looking time-series as shown in Fig.~\ref{fig:dcm_imputation}. Therefore, we looked for a simpler, more informed imputation strategy. The Imputation strategy we chose for our final data pipeline works as follows:
%
(1)  Compute a rolling-window median of the data,
(2)  Replace NaNs with the median,
(3)  Do a backward fill to deal with NaNs at the start, and
(4)  Do a forward fill to deal with NaNs at the end.

We compared the KNNImputer to our simple imputation strategy by looking at the
mean-squared error on the validation dataset with artificially
introduced missing values:
\begin{table}[h!]
    \centering
    \begin{tabular}{lr}
    KNNImputer(n\_neighbors=20, weights=''uniform'') & 0.00230 \\
    Our rolling window approach & 0.00021 \\
    \end{tabular}
\end{table}

This observation justified the usage of our simpler rolling-window
approach, which we used for our experiments.

\begin{figure}
    \centering
    \includegraphics[width=\linewidth]{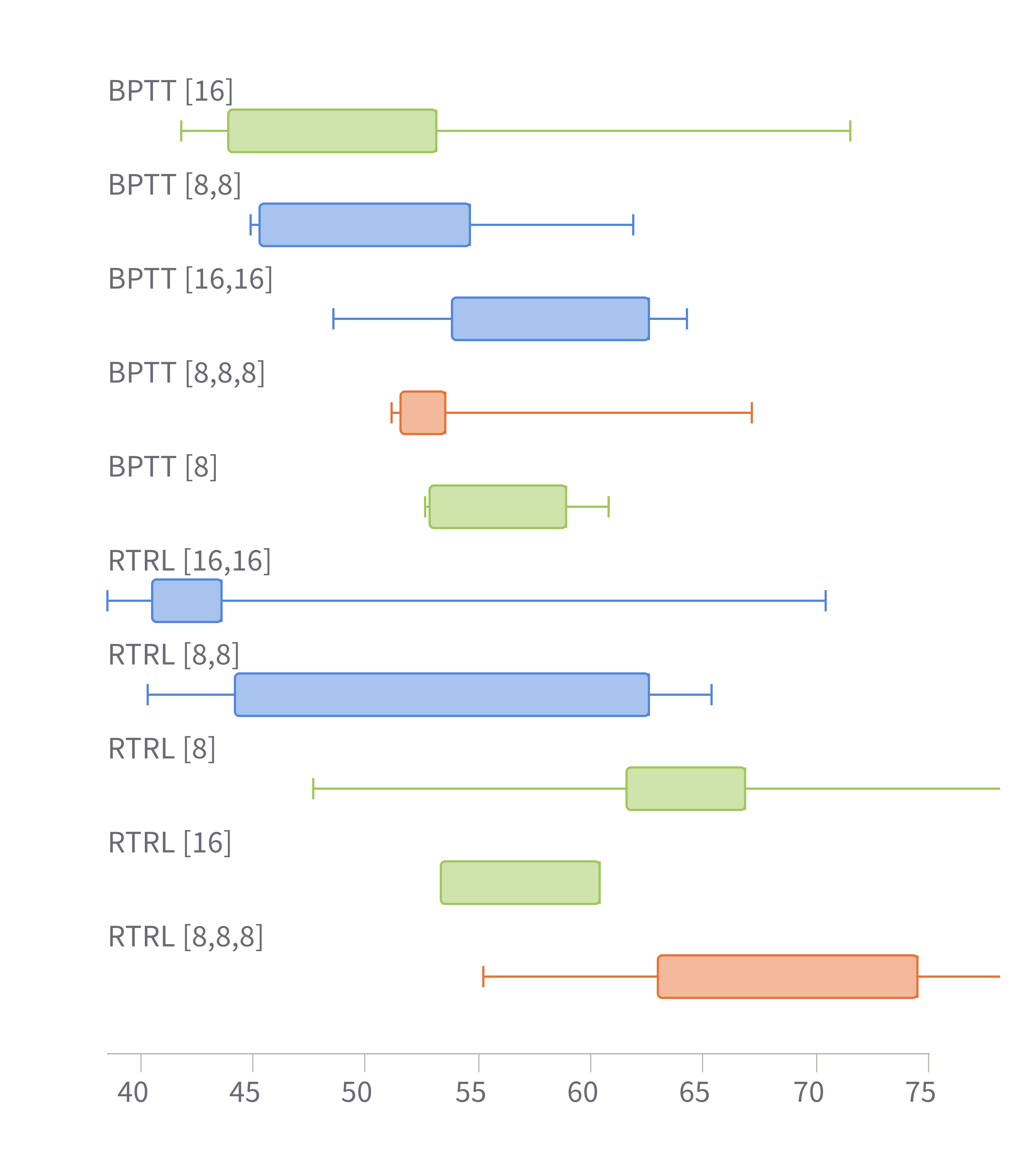}
    \caption{Boxplot of evaluation losses for 5 runs each with different layer sizes, trained with RTRL or BPTT. A single layer with 16 neurons was best for BPTT while a two-layered architecture with 16 neurons had the lowest best-case prediction error for RTRL.}
    \label{fig:box_layers}
\end{figure}

\section{EXPERIMENTS}
We built a carbon emission LRU prediction model, and trained it on the preprocessed dataset, by using the pipeline described above. The model was trained to predict carbon emissions from readily available data, such as engine RPM and outside temperature. Training models from scratch using both backpropagation and RTRL, we conduct a hyperparameter search and examine the results. Then, we built a carbon emission prediction model with the best configuration found, and pre-train it using backpropagation. Finally, we validated the best trained model while fine-tuning the model using RTRL. Our results show that the prediction error can be consistently decreased during inference using our method.

If not stated otherwise, the networks were trained using the Adam optimizer \cite{kingma2017} for 100 thousand steps and with a batch size of 256, learning rate of $10^{-3}$, and 0.5 gradient norm clipping. The loss function to optimize was the Huber loss of ground truth versus predicted values. \cite{huber1964}

\subsection{RNN hyperparameter Search}

\begin{figure}
    \centering
    \includegraphics[width=\linewidth]{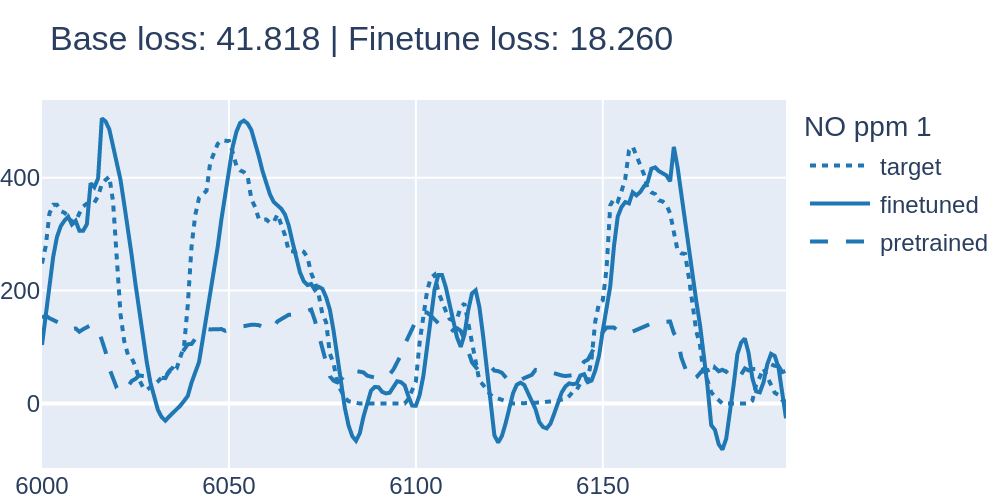}
    \caption{Prediction of NO ppm with and without fine-tuning of a model pre-trained using BPTT. Regularization strength was set to 0.01.}
    \label{fig:finetuning_good_1}
\end{figure}

\begin{figure}
    \centering
    \includegraphics[width=\linewidth]{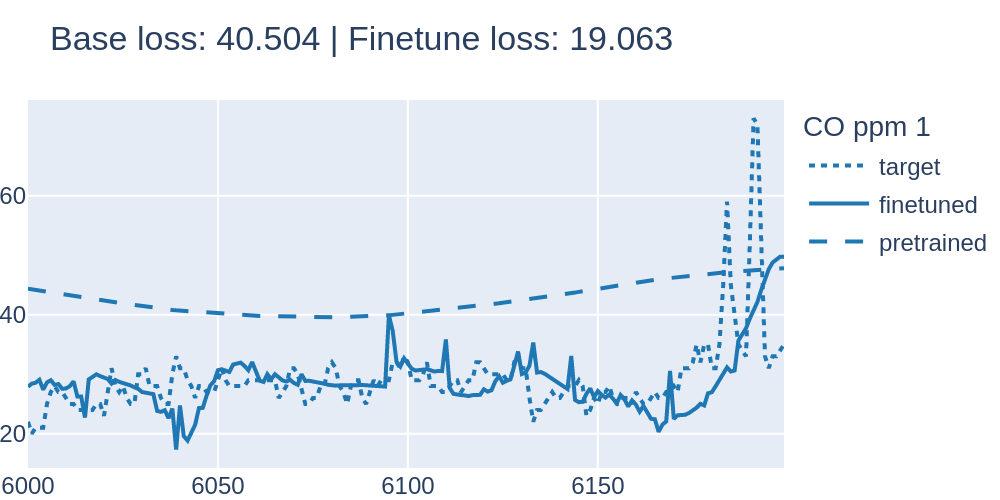}
    \caption{Prediction of CO ppm with and without fine-tuning of a model pre-trained using RTRL. No regularization yielded the best result.}
    \label{fig:finetuning_good_2}
\end{figure}

\begin{figure*}
    \centering
    \includegraphics[width=.25\linewidth]{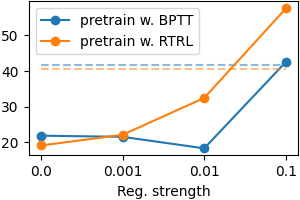}
    \includegraphics[width=0.74\linewidth]{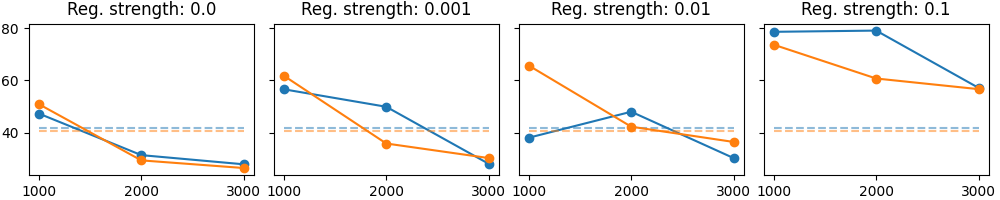}
    \caption{The plots show the total prediction error on the validation sequence when fine-tuning during inference. Dashed lines show the error of the pre-trained model without tuning. Left: Fine-tuning throughout the entire validation sequence for different values of regularization strength. Note that the X-axis is categorical to accommodate both log-scale and zero. Right: Fine-tuning just at the start. X-axis shows the number of fine-tuning steps. While beneficial for BPTT when fine-tuning all the way, regularization is worsening results when only updating parameters at the beginning.}
    
    \label{fig:finetune_until}
\end{figure*}
\begin{figure*}
    \centering
    \includegraphics[width=.43\linewidth]{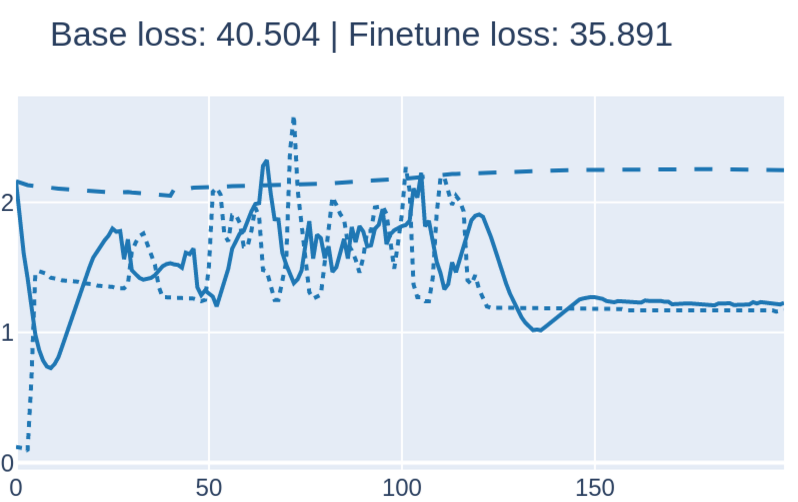}
    \includegraphics[width=.54\linewidth]{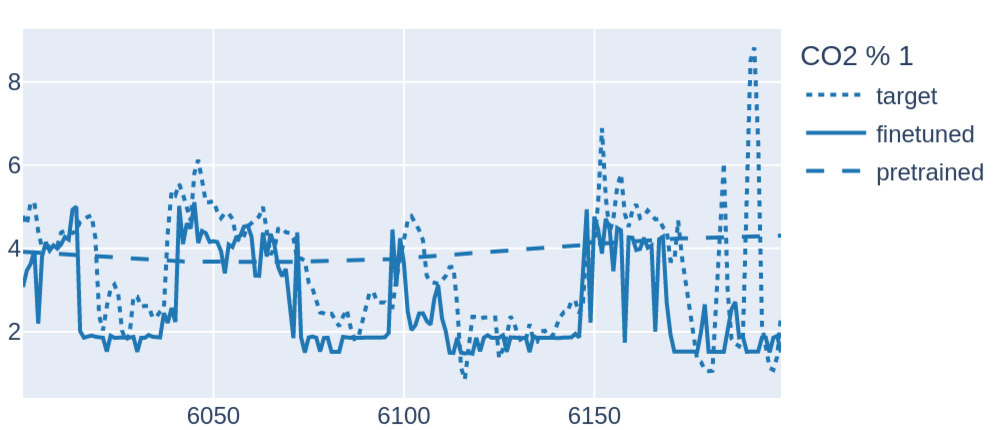}
    \caption{Prediction of CO2\% with a model that was pre-trained using RTRL when only fine-tuning for the initial 2000 steps. Regularization strength was 0.001. Left shows the adjustments manifesting at the start of the sequence. Right shows a sustained improvement of the prediction 4000 steps later.}
    \label{fig:finetuning_bad}
\end{figure*}
We started by pre-training LRU models to predict the carbon emissions. For comparison, we tested both BPTT and RTRL for training from scratch. To combat overfitting and reduce inference time on embedded hardware, we specifically looked into using small architectures. The layer sizes we tested are (8,), (16,), (8, 8), (16, 16), and (8, 8, 8). We also tuned the learning rate with values $10^{-2}$, $10^{-3}$ and $10^{-4}$, and gradient clipping where we tried $0.5$, $1.0$ and no clipping.

Fig.~\ref{fig:box_layers} shows the best validation loss for 5 runs each. We found
that (16,) resulted in the best fit for BPTT while for RTRL (16, 16) achieved the best validation loss. Our hyperparameter search further showed that a learning rate of $10^{-3}$ and clipping the gradients to have a maximum norm of $0.5$ led to the best results. These values were used for the fine-tuning experiments in the next section.

\subsection{Online Fine-Tuning}

Using RTRL, we can compute gradients online allowing us to fine-tune a pre-trained model during inference. This can be helpful when training on data that do not fully represent the final environment of application. One example might be facing extreme weather conditions that are not present in the training data or a change of driving style during inference.

For this, we pre-trained a model using the method outlined in the previous section. Then we fine-tuned during inference, minimizing the same prediction loss and using the same optimizer, updating the model after each step. This assumes sensor-data availability at all times. To ensure that our fine-tuned model stays close to the pre-trained one, we introduce a regularization term to the loss that penalizes the L2-norm between the pre-trained and current model weights:
\begin{equation}
    R_t = \lambda\ ||\theta_{pre} - \theta_t||_2
\end{equation}
We found that the loss for prediction of the validation set could be reduced by around half when fine-tuning during inference. Fig. \ref{fig:finetuning_good_1} shows a qualitative assessment of the prediction results for NO ppm of a model pre-trained with BPTT. The initial loss for the entire validation sequence was 41.818, whereas with fine-tuning it reduced to 18.260. Similarly, Fig. \ref{fig:finetuning_good_2} shows the prediction of CO ppm of a model pre-trained with RTRL. In both cases, the pre-trained model fails to accurately follow the trend whereas the fine-tuned model matches the emission data almost perfectly.

\subsubsection{Ablation Study}

In order to assess the effect of our regularization method, we conducted an ablation study by fine-tuning the same models with varying regularization strengths of zero, 0.001, 0.01, and 0.1. The results are shown in Fig. \ref{fig:finetune_until} left: Using an intermediate regularization strength was beneficial for models pre-trained with BPTT, but not RTRL. This observation was surprising and further investigation is needed to find an explanation for this discrepancy.

Finally, we investigated the effect of fine-tuning only at the beginning of the validation sequence. To this end, we froze the parameters after training throughout the initial 1000, 2000, or 3000 steps of the validation sequence. Fig. \ref{fig:finetune_until} right shows the results: In almost all instances, a minimum of 2000 steps of training is necessary in order to achieve an improvement of the prediction. When fine-tuning for just a small number of steps, the prediction error increased even. Prediction results for a model that was fine-tuned in this manner for 2000 steps are presented in Fig. \ref{fig:finetuning_bad}. The left panel shows the initial phase of fine-tuning with a pronounced departure of the fine-tuned prediction from the pre-trained one. The right panel shows how the improvements are being sustained even after a large number of steps has passed.

\section{CONCLUSIONS}

We introduced the idea of fine-tuning structured state-space models with real-time recurrent learning and empirically demonstrated the validity of the approach using a carbon emission dataset recorded from embedded hardware.

The novel technique is not limited to continuous regression tasks as conducted in this work, but also applicable to a wide range of sequence modeling problems such as machine translation with large language models, representation learning in robotic manipulation, or state-value estimation in reinforcement learning with partially observable environments \cite{lemmel2025}.

Contrary to our initial intuition, pre-training with RTRL does not give an advantage when fine-tuning using the algorithm. Our experimental results showed that models that were pre-trained with BPTT can later be fine-tuned just as well using RTRL. Furthermore, using L2-regularization for constraining the model parameters to stay close to the pre-trained ones did only work for models pre-trained with BPTT and when fine-tuning for the entire prediction horizon. A possible improvement that should be explored in future work is the implementation of continual learning methods such as elastic weight consolidation \cite{kirkpatrick2017} or synaptic intelligence \cite{zenke2017}.




\section*{ACKNOWLEDGMENT}

This work was funded by Austrian Research Promotion Agency (FFG) Project
grant No. FO999899799.

\addtolength{\textheight}{-1cm}   
                                  
\printbibliography

@techreport{BlellochTR90,
  title = {Prefix Sums and Their Applications},
  author = {Blelloch, Guy E.},
  year = {1990},
  month = nov,
  number = {CMU-CS-90-190},
  institution = {School of Computer Science, Carnegie Mellon University}
}

@article{fujita2012,
  title = {Comparison of the {{MOVES2010a}}, {{MOBILE6}}.2, and {{EMFAC2007}} Mobile Source Emission Models with on-Road Traffic Tunnel and Remote Sensing Measurements},
  author = {Fujita, Eric M. and Campbell, David E. and Zielinska, Barbara and Chow, Judith C. and Lindhjem, Christian E. and DenBleyker, Allison and Bishop, Gary A. and Schuchmann, Brent G. and Stedman, Donald H. and Lawson, Douglas R.},
  year = {2012},
  month = oct,
  journal = {Journal of the Air \& Waste Management Association},
  volume = {62},
  number = {10},
  pages = {1134--1149},
  publisher = {Taylor \& Francis},
  issn = {1096-2247},
  doi = {10.1080/10962247.2012.699016},
}

@inproceedings{gu2021a,
  title = {Efficiently {{Modeling Long Sequences}} with {{Structured State Spaces}}},
  booktitle = {International {{Conference}} on {{Learning Representations}}},
  author = {Gu, Albert and Goel, Karan and Re, Christopher},
  year = {2021},
  month = oct,
  langid = {english}
}

@article{huber1964,
  title = {Robust {{Estimation}} of a {{Location Parameter}}},
  author = {Huber, Peter J.},
  year = {1964},
  month = mar,
  journal = {The Annals of Mathematical Statistics},
  volume = {35},
  number = {1},
  pages = {73--101},
  publisher = {Institute of Mathematical Statistics},
  issn = {0003-4851, 2168-8990},
  doi = {10.1214/aoms/1177703732}
}

@article{kingma2017,
  title = {Adam: {{A Method}} for {{Stochastic Optimization}}},
  shorttitle = {Adam},
  author = {Kingma, Diederik P. and Ba, Jimmy},
  year = {2017},
  month = jan,
  journal = {arXiv:1412.6980 [cs]},
  eprint = {1412.6980},
  primaryclass = {cs},
  archiveprefix = {arXiv},
}

@article{kirkpatrick2017,
  title = {Overcoming Catastrophic Forgetting in Neural Networks},
  author = {Kirkpatrick, James and Pascanu, Razvan and Rabinowitz, Neil and Veness, Joel and Desjardins, Guillaume and Rusu, Andrei A. and Milan, Kieran and Quan, John and Ramalho, Tiago and {Grabska-Barwinska}, Agnieszka and Hassabis, Demis and Clopath, Claudia and Kumaran, Dharshan and Hadsell, Raia},
  year = {2017},
  month = mar,
  journal = {Proceedings of the National Academy of Sciences},
  volume = {114},
  number = {13},
  eprint = {1612.00796},
  pages = {3521--3526},
  publisher = {Proceedings of the National Academy of Sciences},
  doi = {10.1073/pnas.1611835114},
  archiveprefix = {arXiv}
}

@article{lemmel2025,
  title = {Real-{{Time Recurrent Reinforcement Learning}}},
  author = {Lemmel, Julian and Grosu, Radu},
  year = {2025},
  month = apr,
  journal = {Proceedings of the AAAI Conference on Artificial Intelligence},
  volume = {39},
  number = {17},
  pages = {18189--18197},
  issn = {2374-3468},
  doi = {10.1609/aaai.v39i17.34001},
  copyright = {All rights reserved},
  langid = {english},
}

@inproceedings{mahendra2024,
  title = {Predicting {{Future Aircraft Emissions}} in {{Indonesia}}: {{A Comparative Analysis}} Using {{LSTM}} and {{GRU Methods}}},
  shorttitle = {Predicting {{Future Aircraft Emissions}} in {{Indonesia}}},
  booktitle = {Proceedings of the 2nd {{International Conference}} on {{Aviation Industry}}, {{Education}}, and {{Regulation}}, {{AVINER}} 2023, 8-9 {{November}} 2023, {{Jakarta}}, {{Indonesia}}},
  author = {Mahendra, Fhilipus and S, Reici and Hannim, Sheyla and Arifin, Mufti},
  year = {2024},
  month = may,
  isbn = {978-1-63190-467-7}
}

@inproceedings{murti2019,
  title = {K-{{Nearest Neighbor}} ({{K-NN}}) Based {{Missing Data Imputation}}},
  booktitle = {2019 5th {{International Conference}} on {{Science}} in {{Information Technology}} ({{ICSITech}})},
  author = {Murti, Della Murbarani Prawidya and Pujianto, Utomo and Wibawa, Aji Prasetya and Akbar, Muhammad Iqbal},
  year = {2019},
  month = oct,
  pages = {83--88},
  doi = {10.1109/ICSITech46713.2019.8987530},
}

@inproceedings{ntziachristos2009,
  title = {{{COPERT}}: {{A European Road Transport Emission Inventory Model}}},
  shorttitle = {{{COPERT}}},
  booktitle = {Information {{Technologies}} in {{Environmental Engineering}}},
  author = {Ntziachristos, Leonidas and Gkatzoflias, Dimitrios and Kouridis, Chariton and Samaras, Zissis},
  editor = {Athanasiadis, Ioannis N. and Rizzoli, Andrea E. and Mitkas, Pericles A. and G{\'o}mez, Jorge Marx},
  year = {2009},
  pages = {491--504},
  publisher = {Springer},
  address = {Berlin, Heidelberg},
  doi = {10.1007/978-3-540-88351-7_37},
  isbn = {978-3-540-88351-7},
  langid = {english},
}

@inproceedings{orvieto2023,
  title = {Resurrecting Recurrent Neural Networks for Long Sequences},
  booktitle = {Proceedings of the 40th {{International Conference}} on {{Machine Learning}}},
  author = {Orvieto, Antonio and Smith, Samuel L and Gu, Albert and Fernando, Anushan and Gulcehre, Caglar and Pascanu, Razvan and De, Soham},
  year = {2023},
  month = jul,
  series = {{{ICML}}'23},
  volume = {202},
  pages = {26670--26698},
  publisher = {JMLR.org},
  address = {Honolulu, Hawaii, USA}
}

@article{williams1989,
  title = {A {{Learning Algorithm}} for {{Continually Running Fully Recurrent Neural Networks}}},
  author = {Williams, Ronald J. and Zipser, David},
  year = {1989},
  month = jun,
  journal = {Neural Computation},
  volume = {1},
  number = {2},
  pages = {270--280},
  issn = {0899-7667},
  doi = {10.1162/neco.1989.1.2.270},
}

@article{xu2021,
  title = {Spatiotemporal {{Graph Convolution Multifusion Network}} for {{Urban Vehicle Emission Prediction}}},
  author = {Xu, Zhenyi and Kang, Yu and Cao, Yang and Li, Zhijun},
  year = {2021},
  month = aug,
  journal = {IEEE Transactions on Neural Networks and Learning Systems},
  volume = {32},
  number = {8},
  pages = {3342--3354},
  issn = {2162-2388},
  doi = {10.1109/TNNLS.2020.3008702},
}

@misc{zenke2017,
  title = {Continual {{Learning Through Synaptic Intelligence}}},
  author = {Zenke, Friedemann and Poole, Ben and Ganguli, Surya},
  year = {2017},
  month = jun,
  number = {arXiv:1703.04200},
  eprint = {1703.04200},
  publisher = {arXiv},
  doi = {10.48550/arXiv.1703.04200},
  archiveprefix = {arXiv}
}

@article{zhang2018,
  title = {Vehicle {{Emission Forecasting Based}} on {{Wavelet Transform}} and {{Long Short-Term Memory Network}}},
  author = {Zhang, Qiang and Li, Feng and Long, Fei and Ling, Qiang},
  year = {2018},
  journal = {IEEE Access},
  volume = {6},
  pages = {56984--56994},
  issn = {2169-3536},
  doi = {10.1109/ACCESS.2018.2874068}
}

@article{zhong2024,
  title = {Models for Predicting Vehicle Emissions: {{A}} Comprehensive Review},
  shorttitle = {Models for Predicting Vehicle Emissions},
  author = {Zhong, Hui and Chen, Kehua and Liu, Chenxi and Zhu, Meixin and Ke, Ruimin},
  year = {2024},
  month = may,
  journal = {Science of The Total Environment},
  volume = {923},
  pages = {171324},
  issn = {0048-9697},
  doi = {10.1016/j.scitotenv.2024.171324}
}

@inproceedings{zucchet2023,
  title = {Online Learning of Long-Range Dependencies},
  booktitle = {Thirty-Seventh {{Conference}} on {{Neural Information Processing Systems}}},
  author = {Zucchet, Nicolas and Meier, Robert and Schug, Simon and Mujika, Asier and Sacramento, Joao},
  year = {2023},
  month = nov,
  langid = {english}
}

\end{document}